\documentclass[nofootinbib,reprint,longbibliography]{revtex4-1}
\usepackage{xr}
\usepackage{color}
\definecolor{light-gray}{gray}{0.95}

\newcommand{\shadedbox}[1]{\colorbox{light-gray}{$\displaystyle #1$}}

\usepackage{amsmath,amssymb}
\usepackage{graphicx}
\usepackage[abs]{overpic}
\newcommand\rr{\mathbf{r}}
\newcommand\pp{\mathbf{p}}

\newcommand\zhat{\mathbf{\hat{z}}}
\newcommand\phat{\mathbf{\hat{\pp}}}
\newcommand\rhat{\mathbf{\hat{r}}}

\newcommand\aone{\mathbf{a}_1^\omega}
\newcommand\eone{\mathbf{e}_1^\omega}
\newcommand\tone{\mathbf{t}_1^\omega}

\newcommand\Jrt{\mathbf{J}(\rr,t)}
\newcommand\Jomegagen{\mathbf{J}_\omega(\pp)}
\newcommand\Jomega{\mathring{\mathbf{J}}_\omega(\phat)}

\newcommand\rhoomega{\mathring{\rho}_\omega(\phat)}
\newcommand\Aomega{\mathring{\mathbf{A}}(\phat)}
\newcommand\phiomega{\mathring{\phi}(\phat)}

\newcommand\Jomegar{{\mathbf{J}}_\omega(\rr)}

\newcommand\Xjm{\mathbf{X}_{jm}}
\newcommand\Zjm{\mathbf{Z}_{jm}}
\newcommand\Wjm{\mathbf{W}_{jm}}
\newcommand\Qjm{\mathbf{Q}_{jm}}

\newcommand\Yjlm{\mathbf{Y}_{j,l,m}}
\newcommand\Yjjpm{\mathbf{Y}_{j,j+1,m}}
\newcommand\Yjjm{\mathbf{Y}_{j,j,m}}
\newcommand\Yjjmm{\mathbf{Y}_{j,j-1,m}}

\newcommand\tauj{\tau_{j,j\pm1,m}^\omega}
\newcommand\Sj{{\textstyle\sqrt{\frac{j}{2j+1}}}}
\newcommand\Sjp{{\textstyle\sqrt{\frac{j+1}{2j+1}}}}

\newcommand\ajm{a_{jm}^\omega}
\newcommand\bjm{b_{jm}^\omega}
\newcommand\cjm{c_{jm}^\omega}
\newcommand\qjm{q_{jm}^\omega}

\newcommand\mbar{\overline{m}}
\newcommand\lbar{\bar{l}}

\newcommand\intdOmegap{\int d \phat\text{ }}

\newcommand\intdr{\int d^3\rr\text{ }}

\newcommand\Aomegar{\mathbf{A}_\omega(\rr)}

\newcommand\aoo{a_{11}^\omega}

\newcommand\aom{a_{1-1}^\omega}

\newcommand\aoz{a_{10}^\omega}

\usepackage{color}
\usepackage{bm}

\definecolor{light-gray}{gray}{0.95}

\begin{document}
\title{On the dynamic toroidal multipoles from localized electric current distributions}
\author{Ivan \surname{Fernandez-Corbaton}$^{1}$$^{,*}$}
\author{Stefan Nanz$^{2}$}
\author{Carsten Rockstuhl$^{1,2}$}
\affiliation{$^{1}$Institute of Nanotechnology, Karlsruhe Institute of Technology, 76021 Karlsruhe, Germany}
\affiliation{$^{2}$Institut f\"ur Theoretische Festk\"orperphysik, Karlsruhe Institute of Technology, 76131 Karlsruhe, Germany}
\address{Correspondance to: ivan.fernandez-corbaton@kit.edu}
\begin{abstract}
We analyze the dynamic toroidal multipoles and prove that they do not have an independent physical meaning with respect to their interaction with electromagnetic waves. We analytically show how the split into electric and toroidal parts causes the appearance of non-radiative components in each of the two parts. These non-radiative components, which cancel each other when both parts are summed, preclude the separate determination of each part by means of measurements of the radiation from the source or of its coupling to external electromagnetic waves. In other words, there is no toroidal radiation or independent toroidal electromagnetic coupling. The formal meaning of the toroidal multipoles is clear in our derivations. They are the higher order terms of an expansion of the multipolar coefficients of electric parity with respect to the electromagnetic size of the source.
\end{abstract}
 \maketitle
 \section*{Introduction, summary and outline}
After the introduction of toroidal multipoles in the static case \cite{Zeldovich1958}, the dynamic toroidal multipoles were presented as a new independent multipole family that had been previously ignored \cite{Dubovik1974,Dubovik1990}.  The new family was derived from the split of the transverse multipoles of electric parity into two parts. These parts are often referred to as electric and toroidal. The dynamic toroidal multipoles have also been analyzed in Refs. \onlinecite{Afanasiev1994,Radescu2002}, and are currently being considered in the areas of metamaterials, plasmonics, and nanophotonics \cite{Marinov2007, Kaelberer2010, Zheng2012, Dong2012,Ogut2012,Huang2012, Fan2013, Fedotov2013,Savinov2014,Liu2014, Basharin2015,Liu2015,Liu2015b,Papasimakis2016}.

In this article, we analyze the split between the dynamic electric and toroidal multipoles of a localized source distribution. We find that the two parts cannot be separately determined by measuring the electromagnetic fields produced by the source outside the source region. The two parts can also not be separately determined by measuring the coupling between the source and externally incident electromagnetic waves. The result applies to both near and far field situations, and implies that there is no independent coupling of toroidal character between the sources and the electromagnetic field, and that there is no radiation of pure toroidal character. The toroidal multipoles can hence not be considered an independent family. Rather, our analysis makes clear that the electric and toroidal parts correspond, respectively, to the lowest and the higher order terms of the expansion of the exact multipolar coefficients of electric parity with respect to the electromagnetic size of the source. In this respect, our article contains analytical expressions that inherently contain all correction orders: They are exact for any source size. 

Our results do not question the usefulness of considering the higher order terms. For example, some experimental results can only be explained by adding the toroidal dipole contributions to the lowest order terms (see Fig. 3c in Ref. \onlinecite{Kaelberer2010}, and Fig. 4 in Ref. \onlinecite{Basharin2015}). The same is true for non-radiating spherical configurations (see Fig. 2 in Ref. \onlinecite{Miroshnichenko2014}). These cases are examples where the toroidal terms are the dominant dipolar terms. 

In the following, we first outline the mathematical setting in which we carry out the analysis. We then highlight the direct connections between our momentum space approach and the physics of multipolar couplings between sources and fields. On the one hand, three kinds of degrees of freedom are needed to describe the sources. Two of them are transverse and have opposite (electric vs. magnetic) parity, and the other one is longitudinal. On the other hand, only the two transverse degrees of freedom are needed to describe the electromagnetic fields produced by the sources outside the source region, where the field produced by the longitudinal degrees of freedom is identically zero. 

Then, we analyze the physical meaning of the toroidal multipoles. Our analysis starts from recently obtained exact analytical expressions for the transverse multipolar coefficients of electric parity. These expressions are valid for any source size \cite{FerCor2015b}. We take the dipolar case as the guiding example, make the small source approximation keeping the two lowest orders, and show that: The lowest order term is the well know approximate expression for the electric dipole of electromagnetically small sources; the term of second lowest order is the approximate expression for the toroidal dipole of electromagnetically small sources [Eq. (2.11) in Ref. \onlinecite{Dubovik1990}]. We then analyze the original split \cite{Dubovik1974,Dubovik1990} in the general multipolar case, and establish that both electric and toroidal parts contain non-radiative components, which appear due to the separation of terms of different order in the size of the source. These non-radiative components, which cancel each other when both parts are summed, preclude the separate determination of each part by measuring the radiation from the source outside the source region, or by measuring the coupling between the source and externally incident electromagnetic waves. Afterwards, we analyze a different class of splits where the two parts are free of this kind of non-radiative components. We find that, in this case, what precludes their independent measurement is the absence of longitudinal radiation. We show in App. D that this is the same reason that precludes the independent measurement of the recently proposed re-definition of toroidal multipoles (Box 2 in Ref. \onlinecite{Papasimakis2016}). 

Finally, we use the obtained insights to clarify some statements that are often found in the literature.

\section*{Mathematical setting}\label{sec2}
We start with an electric current density distribution $\Jrt$ embedded in an isotropic homogeneous medium characterized by real valued permittivity $\epsilon$ and permeability $\mu$. We assume that $\Jrt$ is confined in space so that $\Jrt=0$ for $|\rr|>R$. We consider its energy-momentum Fourier representation 
	\begin{eqnarray}
	\label{eq2:first}
		&&\Jrt=\mathcal{R}\left[\int_{0^{+}}^{\infty}\frac{d\omega}{\sqrt{2\pi}}\exp(-i\omega t)\mathbf{J}_\omega(\rr)\right],\\
			   &&=\mathcal{R}\left[\int_{0^{+}}^{\infty}\frac{d\omega}{\sqrt{2\pi}}\exp(-i\omega t)\int \frac{d^3\pp}{\sqrt{(2\pi)^3}}\ \Jomegagen\exp(i\pp\cdot\rr)\right],\nonumber
	\end{eqnarray}
	and treat each frequency $\omega$ separately. The lower limit of the integral in $d\omega$ excludes the static case $\omega=0$, which we do not treat in this paper. This formal setting is independent of the cause of the current. For example, in a scattering situation the current is induced in the scatterer by an external field.

 \begin{figure}[ht!]
	\begin{overpic}[width=0.5\textwidth,unit=1mm]{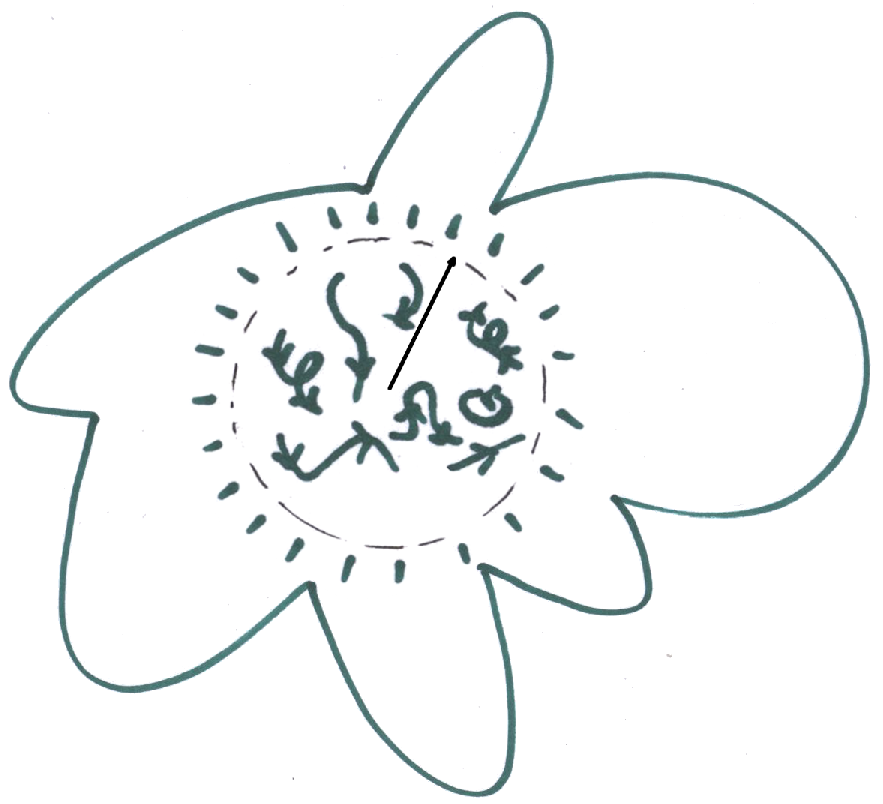}
\put(43.5,45){\large$R$}
\put(56,54){\large$\left[\mathbf{E}_\omega(\rr),\mathbf{H}_\omega(\rr)\right]$}
\put(64,48){\large$|\rr|>R$}
\put(36,29.8){\large$\mathbf{J}_\omega(\rr)$}

	\end{overpic}
	\caption{\label{fig:regions} The monochromatic electric current density distribution $\mathbf{J}_\omega(\rr)$ is confined to a sphere of radius $R$: The {\em source region}. In this article, we consider the electromagnetic fields produced by the source outside its region, i.e. $\left[\mathbf{E}_\omega(\rr),\mathbf{H}_\omega(\rr)\right]$ for $|\rr|>R$. These fields are completely determined by the Fourier components of the current density which meet $|\pp|=\omega/c$ (see Fig. \ref{fig:shell}). The other components do not produce fields outside the source region. This is a general result that is also valid when the confining volume defining the source region is not spherical (see text).}
\end{figure}
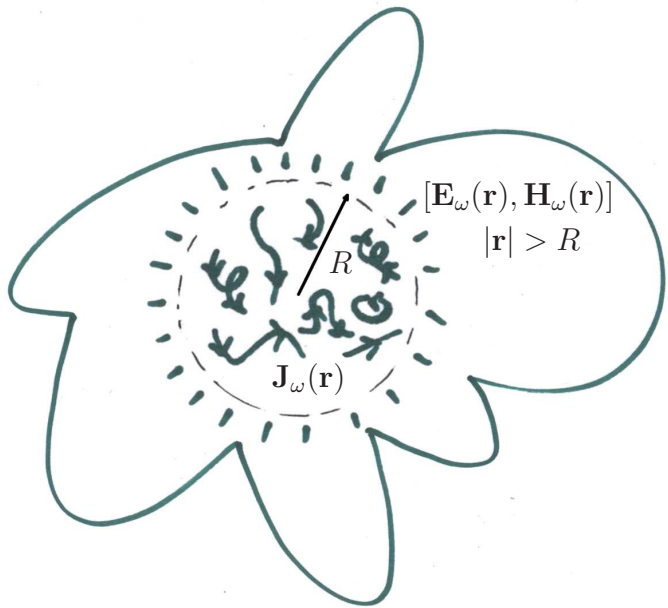

The electromagnetic fields radiated by the source at each frequency $\omega$ are solely determined by the part of $\Jomegagen$ in the domain that satisfies $|\pp|=\omega\sqrt{\epsilon\mu}=\omega/c$. This theorem has been proven in Refs. \onlinecite{Devaney1973,Devaney1974}. The authors showed that this part of the current completely determines the electromagnetic field generated by the source at any point $\rr$ such that $|\rr|>R$, which, in particular, includes near fields. This result has been extended beyond spheres to confining volumes of any smooth enough shape (see Chap. 9 in Ref. \onlinecite{Colton2012}): The part of $\Jomegagen$ in the domain that satisfies $|\pp|=\omega\sqrt{\epsilon\mu}=\omega/c$ determines the electromagnetic field generated by the source at any point {\em outside the source region}, for a wide class of shapes of such source region. Figure \ref{fig:regions} depicts a spatially confined monochromatic source distribution and the fields that it generates outside the source region. 

\begin{figure}[ht!]
	\def\ASYprefix{}
\newbox\ASYbox
\newdimen\ASYdimen
\long\def\ASYbase#1#2{\leavevmode\setbox\ASYbox=\hbox{#1}\setbox\ASYbox=\hbox{#2}\lower\ASYdimen\box\ASYbox}
\long\def\ASYaligned(#1,#2)(#3,#4)#5#6#7{\leavevmode\setbox\ASYbox=\hbox{#7}\setbox\ASYbox\hbox{\ASYdimen=\ht\ASYbox\advance\ASYdimen by\dp\ASYbox\kern#3\wd\ASYbox\raise#4\ASYdimen\box\ASYbox}\setbox\ASYbox=\hbox{#5\wd\ASYbox 0pt\dp\ASYbox 0pt\ht\ASYbox 0pt\box\ASYbox#6}\hbox to 0pt{\kern#1pt\raise#2pt\box\ASYbox\hss}}\long\def\ASYalignT(#1,#2)(#3,#4)#5#6{\ASYaligned(#1,#2)(#3,#4){
}{
}{#6}}
\long\def\ASYalign(#1,#2)(#3,#4)#5{\ASYaligned(#1,#2)(#3,#4){}{}{#5}}
\def\ASYraw#1{
currentpoint currentpoint translate matrix currentmatrix
100 12 div -100 12 div scale
#1
setmatrix neg exch neg exch translate}
 	\begin{overpic}[width=0.5\textwidth,unit=1mm]{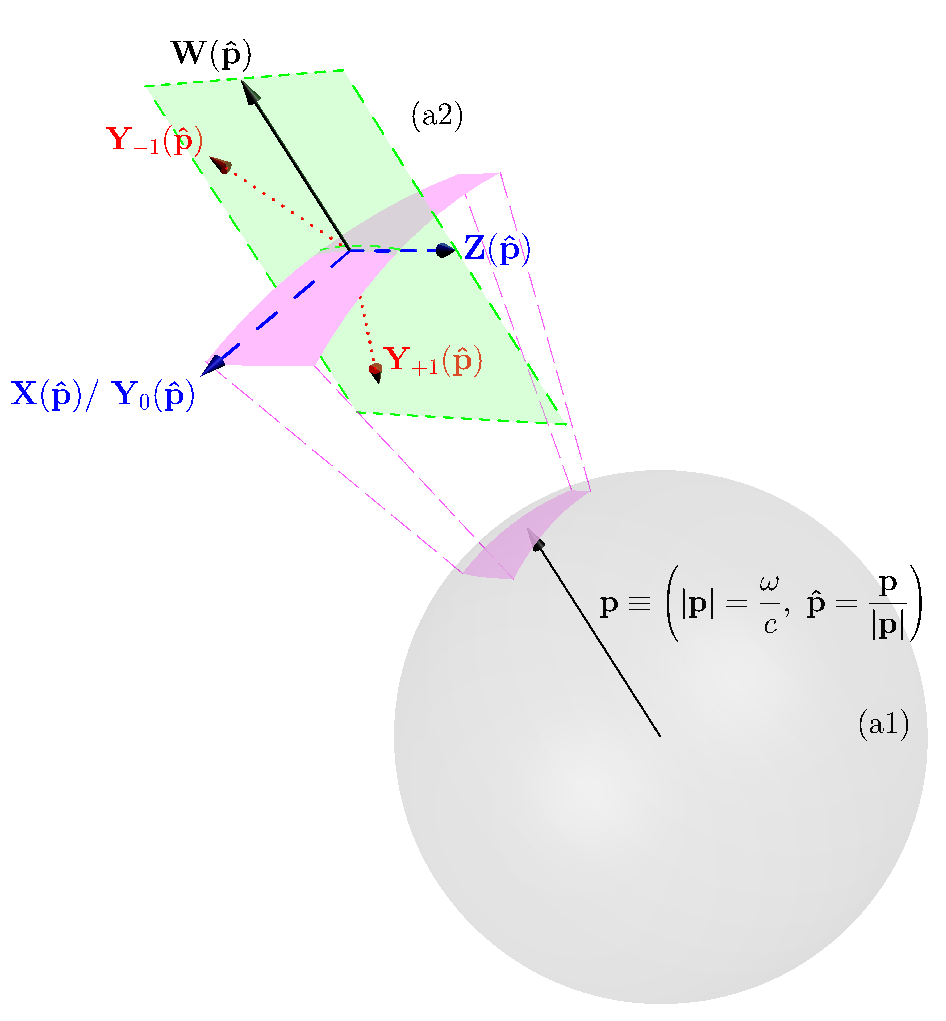}
	\put(0,0){\includegraphics[scale=0.5]{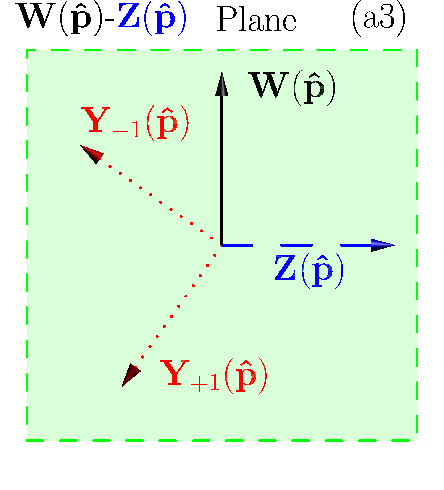}}
		\end{overpic}
	\caption{\label{fig:shell} Only the transverse components of the current density $\mathbf{J}_\omega(\pp)$ with $\pp=\frac{\omega}{c}\phat$ produce electromagnetic fields outside the source region (see Fig. \ref{fig:regions}). (a1) Spherical shell of radius $|\pp|=\frac{\omega}{c}$ in momentum space. The (a2) region depicts two different orthonormal bases for vector functions defined on the shell. The $\{\Xjm(\phat),\Zjm(\phat),\Wjm(\phat)\}$ and the $\{\Yjjm(\phat),\Yjjpm(\phat),\Yjjmm(\phat)\}$. The $(j,m)$ indexes are omitted in the figure, and the subscript in the $\mathbf{Y}$'s is the difference between their second and first indexes. The two bases are related to each other as written in Eq. (\ref{eq2:Y}). With respect to their polarization vectors: $\Xjm(\phat)$, $\Zjm(\phat)$, and $\Yjjm(\phat)$ are orthogonal (transverse) to the momentum vector $\pp$, $\Wjm(\phat)$ is parallel (longitudinal) to $\pp$, and $\Yjjpm(\phat)$ and $\Yjjmm(\phat)$ lay in the $\Wjm(\phat)-\Zjm(\phat)$ plane and are neither transverse nor longitudinal, as seen in (a3). The transverse vectors are represented by dashed blue arrows, the longitudinal ones by solid black arrows and the ones of mixed character by dotted red arrows. For the figure, we took $j=1$ in Eq. (\ref{eq2:Y}) for the relationships between $\{\mathbf{Z}(\phat),\mathbf{W}(\phat)\}$ and $\{\mathbf{Y}_{+1}(\phat),\mathbf{Y}_{-1}(\phat)\}$.}
\end{figure}

We denote by $\Jomega$ the components of $\mathbf{J}_\omega(\pp)$ in the spherical shell of radius $|\pp|=\omega/c$ as a function of the solid angle $\phat=\pp/|\pp|$ [see region (a1) in Fig. \ref{fig:shell}]. 

In this domain, the relevant scalar product is $\unexpanded{\langle A | B \rangle = \intdOmegap \mathbf{A}^\dagger(\phat)\mathbf{B}(\phat)}$, where $^{\dagger}$ denotes conjugate transpose, and $\phat$ runs over the entire spherical shell.

We expand $\Jomega$ in an orthonormal basis for vector functions defined in a spherical shell: 
\begin{equation}
	\label{eq2:expand}
	\Jomega=\sum_{jm} \ajm\Zjm(\phat)+\bjm\Xjm(\phat)+\cjm\Wjm(\phat).
\end{equation}

The basis is composed by the three families of vector multipolar functions in momentum space (Sec. B$_I$.3 in Ref. \onlinecite{Cohen1997})

\begin{equation}
	\label{eq2:xzw}
	\begin{split}
		\Xjm(\phat)&=\frac{1}{\sqrt{j(j+1)}} \mathbf{L}Y_{jm}(\phat),\\ 
		\Zjm(\phat)&=i\phat\times\Xjm(\phat),\\ 
		\Wjm(\phat)&=\phat Y_{jm}(\phat),
	\end{split}
\end{equation}
where the $Y_{jm}(\phat)$ are the scalar spherical harmonics and the three components of the vector $\mathbf{L}$ are the angular momentum operators for scalar functions. 

Each of the functions in Eq. (\ref{eq2:xzw}) is an eigenstate of the angular momentum squared $J^2$, the angular momentum along one axis $\mathbf{\hat{\alpha}}$, for which we choose $\mathbf{\hat{\alpha}}=\zhat$, and the parity operator $\Pi$. As depicted in Fig. \ref{fig:shell}, the polarization of $\Xjm(\phat)$ and $\Zjm(\phat)$ is orthogonal (transverse) to $\phat$, and the polarization of $\Wjm(\phat)$ is parallel (longitudinal) to $\phat$ [see region (a2) in Fig. \ref{fig:shell}]. In coordinate space ($\rr$), this distinction corresponds to the distinction between divergence free (transverse) and curl free (longitudinal) vector fields. Table \ref{tab:t1} contains the eigenvalues for the aforementioned three operators and the transverse ($\perp$) or longitudinal ($\parallel$) character of each vector multipolar function.
\begin{table}[h!]
	\begin{ruledtabular}
\begin{center}\begin{tabular}{ccccc}
 & {$J^2$} & {$J_z$} & {$\Pi$}&Polarization\\
	\hline
 $\Xjm(\phat)$ & $j(j+1)$ & $m$ & $(-1)^{j+1}$&$\perp$\\
 $\Zjm(\phat)$ & $j(j+1)$ & $m$ & $(-1)^{j}$&$\perp$\\
 $\Wjm(\phat)$ & $j(j+1)$ & $m$ & $(-1)^{j}$&$\parallel$\\
\end{tabular}\end{center}
	\end{ruledtabular}
\caption{\label{tab:t1}Vector multipolar functions: Polarization character and eigenvalues of $J^2$, $J_z$ and parity $\Pi$. The symbol $\perp$ means transverse polarization. The symbol $\parallel$ means longitudinal polarization. Both $j$ and $m$ are integers, and $m=-j\ldots j$. For $\Xjm(\phat)$ and $\Zjm(\phat)$, $j>0$. For $\Wjm(\phat)$, $j\ge0$.} 
\end{table}

We provide expressions for the $\{\ajm,\bjm,\cjm\}$ in App. A, which we derived in Ref. \onlinecite{FerCor2015b}. 
\subsection*{Multipolar coupling between sources and fields in momentum space}
The decomposition in Eq. (\ref{eq2:expand}) is useful for discussing the physics of multipolar couplings between the sources and the fields. The complex scalars $\{\ajm,\bjm\}$ determine the coupling of the source to the transverse electromagnetic field, a.k.a radiation field, a.k.a electromagnetic waves. The one-to-one relationship between the $\{\ajm,\bjm\}$ coefficients of the source and the multipolar fields of frequency $\omega$ radiated by it is exposed in App. B of Ref. \onlinecite{Blatt1952}, Chap. 9.10 in Ref. \onlinecite{Jackson1998}, and Ref. \onlinecite{Devaney1974}, for example, where the analysis is made in the $\rr$ domain. The $\{\ajm,\bjm\}$ represent independent degrees of freedom in the following sense. Each individual coefficient is connected to the radiation (absorption) of a particular kind of electromagnetic wave: A transverse multipolar field with well defined frequency, angular momentum, and parity. The $\ajm$ are connected to the multipolar waves of electric parity and the $\bjm$ to those of magnetic parity (see Tab. \ref{tab:t1}). The multipolar waves connected to different coefficients are orthogonal to each other. In principle, each individual $\{\ajm,\bjm\}$ can be independently measured through the interaction of the source with the radiation field.  

The $\cjm$ represent the longitudinal degrees of freedom of $\Jomega$, but the longitudinal electric field with $|\pp|=\omega/c$ is zero outside the source region defined in Fig. \ref{fig:regions}. This can be explained in the following way. The longitudinal components of the current ($\cjm$) generate a longitudinal electric field outside the source region defined in Fig. \ref{fig:regions}, but such field is canceled by the longitudinal electric field produced by the associated charge density. This statement follows from the continuity equation, as we show in App. B. The cancellation can also be recognized in \S 13.3 p1875-1877 in Ref. \onlinecite{Morse1953}, and in the expression of the electric field in Eq. (2.39) in Ref. \onlinecite{Radescu2002}, where the two contributions to the longitudinal field cancel exactly. In Ref. \onlinecite{Radescu2002} the longitudinal wave functions are $\mathcal{F}^{(-)}_{jm(\omega/c)}(\rr)$.

The bottom line is that the $\{\ajm,\bjm\}$ contain the necessary and sufficient information to determine both the fields of frequency $\omega$ produced by the sources outside the source region, and the coupling of the sources to external electromagnetic waves of such frequency.

The lack of longitudinal radiation means that, even though there are three kinds of degrees of freedom in the current, $\{\ajm,\bjm,\cjm\}$, only two of them are observable from outside the source region. The powerful theory developed by Gustav Mie for spherical scatterers \cite{Mie1908} is probably the simplest example of this. In this case, the current density is induced in the sphere by an external incident field. The $\{\ajm,\bjm\}$ of this induced current are determined by the Mie coefficients, which can be obtained analytically thanks to the symmetry of the problem. Using the Mie coefficients, the field produced by the induced current can be determined exactly everywhere outside the sphere. This allows the exact calculation of near fields, which is crucial in many applications like for example: Determination of the transition rates of atoms placed near spheres \cite{Chew1987}, determination of the forces and torques caused by evanescent fields onto a sphere near a substrate \cite{Chang1998}, modeling the tips of scanning tunnelling optical microscopes \cite{Barchiesi1993}, and general optical trapping analysis \cite{Dienerowitz2008}. 
\section*{The dynamic toroidal multipoles\label{sec3}}
We will now study the split of the multipolar coefficients of electric parity ($\ajm$) which lead to the appearance of the dynamic toroidal multipoles \cite{Dubovik1974,Dubovik1990}. We use the dipolar case $j=1$ as our guiding example and then generalize the findings to any $j\ge 1$. 

We first show that the well known expressions for the electric and toroidal dipoles of small sources are, respectively, the lowest and second lowest order terms in a series expansion of the exact expression of $a_{1m}^\omega$.

\subsection*{Formal meaning} 
We start from the recently obtained exact expression for the dipolar vector of electric parity in the spherical vector basis $\aone=[\aoo,\aoz,\aom]^T$ (Eq. (\ref{eq:aexact}) in Ref. \onlinecite{FerCor2015b}):
\begin{equation}
	\label{eq2:aexact}
		\begin{split}
			\aone&=-\frac{1}{\pi\sqrt{3}}\intdr  \Jomegar j_0(kr)\\
		&-\frac{1}{2\pi\sqrt{3}}\intdr\left\{3\left[\rhat^{\dagger}\Jomegar\right]\rhat-\Jomegar\right\}j_2(kr),
		\end{split}
\end{equation}
where $k=\omega/c$, and $^\dagger$ means conjugate transpose. Let us now make the small argument approximation to the spherical Bessel functions with terms up to second order [ $j_0(kr)\approx 1-(kr)^2/6$ and $j_2(kr)\approx (kr)^2/15$], and then group the contributions with the same power of $k$.  
\begin{eqnarray}
\label{eq2:eapprox}
&\aone\approx \ \underbrace{-\frac{1}{\pi\sqrt{3}}\intdr \Jomegar}_{\eone}\\
\label{eq2:tapprox}
&\ -\frac{1}{\pi\sqrt{3}} k^2 \underbrace{\intdr\frac{1}{10}\left\{\left[\rr^{\dagger}\Jomegar\right]\rr-2r^2\Jomegar\right\}}_{\tone}.
\end{eqnarray}
We find that the $(k)^0$ term, $\eone$ in Eq. (\ref{eq2:eapprox}), is the well known small source approximation of the electric dipole moment of a current density distribution [Eq. (9.14) in Ref. \onlinecite{Jackson1998}]. The $k^2$ term , $\tone$ in Eq. (\ref{eq2:tapprox}), is the small source approximation of the toroidal dipole moment [Eq. (2.11) in Ref. \onlinecite{Dubovik1990}]. The $\tone$ integral in Eq. (6) contains contributions from the two integrals in Eq. (4): It is the sum of the terms of order $k^2$ coming from both $j_0(kr)$ in the first integral of Eq. (4) and $j_2(kr)$ in the second.

It is now clear that $\tone$ is nothing but the next to leading order term in the electromagnetically small source approximation of the exact expression of $\aone$. All higher order terms are easily obtained from the Taylor series of the spherical Bessel functions in Eq. (\ref{eq2:aexact}). 

The same structure revealed in Eqs. (\ref{eq2:aexact}) to (\ref{eq2:tapprox}) underlies the general case of the electric/toroidal split for $j\ge 1$. In Refs. \onlinecite{Dubovik1974,Dubovik1990}, the lowest order term in $\ajm$, of order $(k)^{j-1}$, is isolated as in Eq. (\ref{eq2:eapprox}). The rest, $\aone-\eone$ in our example, is called {\em toroidal multipole form factor}. Then, the lowest order term of the {\em toroidal multipole form factor}, of order $(k)^{j+1}$ [like our Eq. (\ref{eq2:tapprox})], is called {\em toroidal multipole moment}. For simplicity, we refer to the {\em toroidal multipole form factor} as the {\em toroidal part}. It corresponds to the toroidal multipolar family as defined in Refs. \onlinecite{Dubovik1974,Dubovik1990}. In the dipolar example, the toroidal part ($\aone-\eone$) is equal to $\frac{-k^2}{\pi\sqrt{3}}\tone$ plus all other higher order corrections coming from the terms $(k)^{s>2}$ of the expansion of the spherical Bessel functions in Eq. (\ref{eq2:aexact}).

The formal origin of the toroidal multipoles is recognized in the literature \cite{Afanasiev1994,Liu2014,Miroshnichenko2014,Liu2015b}. We now investigate their physical meaning. 

\subsection*{Physical meaning} 
The splitting of $\ajm$ into two parts \cite{Dubovik1974,Dubovik1990} has a physically relevant consequence. Both parts, often called electric and toroidal, contain Fourier components of the current from outside the $|\pp|=\omega/c$ shell. The $|\pp|\neq\omega/c$ components do not couple to electromagnetic waves \cite{Devaney1974}. For example, it follows from the properties of the Fourier transform that the result of the integral $\int d^3\rr \Jomegar$ in the expression of the electric part in Eq. (\ref{eq2:eapprox}) is proportional to $\mathbf{J}_\omega(\pp=0)$. Since $|\pp|=0\neq\omega/c$, it follows that the toroidal part, $\aone-\eone$, must also contain a contribution proportional to $\mathbf{J}_\omega(\pp=0)$. This is so because $\aone$ does not have any out of shell contribution [Eq. (4.12a) in Ref. \onlinecite{Devaney1974}]: {\em The out of shell contributions in the electric part must be canceled by those in the toroidal part when the two parts are summed}. Since the $|\pp|\neq \omega/c$ components do not couple to the transverse electromagnetic field, it is impossible to determine the toroidal (electric) parts of $\aone$ by measuring the fields produced by the source outside its region, or by externally exciting the source with electromagnetic waves. As previously discussed, this conclusion does not depend on whether the measurement(excitation) occurs in(from) the near, mid, or far field zones, as long as they occur outside the source region. For the sake of discussion, let us now consider the hypothetical case where the first term in the Taylor series of a given component of $\aone$ is zero at a particular frequency. We make no judgment about whether this hypothesis is physically realizable. In this case, the toroidal part would be equal to the said component of $\aone$ and, as such, it would couple to the transverse electromagnetic field. Nevertheless, in this hypothetical case the statement that the toroidal part couples to the field means nothing else than that $\aone$ couples to the field, and it would not amount to separate coupling of the electric and toroidal parts.

The same conclusion applies to any value of $j\ge1$, as we show in App. C. The root cause is the breaking up of the spherical Bessel functions in the split \cite{Dubovik1974,Dubovik1990}. Their lowest order term, proportional to $(kr)^{j-1}$, goes to the electric part and the rest of the terms go to the toroidal part. When unsplit, the spherical Bessel functions in exact integral expressions like Eq. (\ref{eq2:aexact}) or Eq. (\ref{eq2:explicit}) completely reject the $|\pp|\neq \omega/c$ components of $\Jomegar$. The Fourier transforms of spherical Bessel functions correspond to momentum space radial deltas $\delta(|\pp|-\omega/c)$ (see end of Sec. III in Ref. \onlinecite{FerCor2015b}). After the split, the term $(kr)^{j-1}$ by itself does not provide such rejection. This allows $|\pp|\neq \omega/c$ components to ``leak into'' the electric part, and to be forcefully present in the toroidal part as well since the sum of the two parts does not contain any out of shell contribution. 

We have hence established that the electric and toroidal parts cannot be separately determined by measuring the electromagnetic fields produced by the source outside its region, or by measuring its coupling to external electromagnetic waves. Our result implies that, as opposed to electric and magnetic multipoles whose distinct character is given by their parity, there is no electromagnetic field of pure toroidal character. It also implies that toroidal multipoles cannot be independently excited, i.e. there is no independent toroidal coupling. This precludes the existence of selection rules of toroidal character. The two different parities (electric and magnetic), and their linear combinations are necessary and sufficient to characterize electromagnetic transitions (see e.g. Chap. 9.8 p. 436 in Ref. \onlinecite{Jackson1998}, and Chap. XII 2. B in Ref. \onlinecite{Blatt1952}). The lack of longitudinal coupling together with the binary character of the parity operator leave no room for a third option.

While the $\ajm$ have an independent physical meaning, the individual electric and toroidal parts do not have it. The toroidal multipoles cannot be considered an independent family. The same can be said about the isolated electric part, or about any split of the terms in the Taylor series of the spherical Bessel functions in Eqs. (\ref{eq2:explicit}) or Eq. (\ref{eq2:aexact}). Any such split is bound to introduce non-radiative out of shell components in each of its parts.

Since the toroidal multipoles are higher order terms in the approximation of physically meaningful quantities, they can be used to improve analytical models that use only the lowest order terms. The inclusion of the next to leading order term improves the accuracy and prediction ability of such models, as can be seen in Refs. \onlinecite{Kaelberer2010,Huang2012, Fan2013, Savinov2014,Liu2014, Basharin2015,Liu2015,Liu2015b}. In Ref. \onlinecite{Miroshnichenko2014}, and in the context of non-radiating source configurations, the inclusion of the second order term explains a zero in the induced $\aone$, which cannot be explained by the first order term alone (see Fig. 2 in Ref. \onlinecite{Miroshnichenko2014}). The improvement of the models will be maximized by using expressions like Eq. (\ref{eq2:aexact}), Eq. (20) in Ref. \onlinecite{FerCor2015b}, and Eq. (\ref{eq2:explicit}). These expressions inherently contain all correction orders because they are exact for any size of the source.

\subsection*{Splits without out of shell components} 
We have seen that the out of shell non-radiative components appear when splitting $\ajm$ by separating terms of different orders in the size of the source. Nevertheless, we observe that the $\ajm$ in Eqs. (\ref{eq2:explicit}) [$a_{1m}^\omega$ in Eq. (\ref{eq2:aexact})] have two formally distinct contributions, and that each of them contains a spherical Bessel function which guarantees the restriction to $|\pp|=\omega/c$, thereby avoiding the intrusion of the out of shell components. However, we will now show that, due to the absence of longitudinal radiation, the two on-shell contributions cannot be separately determined. 
We consider another orthonormal basis for vector functions on the momentum shell: The vector spherical harmonics $\mathbf{Y}_{j,l,m}(\phat)$. They are related to the $\{\Xjm(\phat),\Zjm(\phat),\Wjm(\phat)\}$ as 
\begin{eqnarray}
\nonumber \Xjm(\phat)&=&\Yjjm(\phat),\\
\nonumber \Zjm(\phat)&=&-\Sjp\Yjjmm(\phat)-\Sj\Yjjpm(\phat),\\
\label{eq2:Y}
\Wjm(\phat)&=&\Sj\Yjjmm(\phat)-\Sjp\Yjjpm(\phat).
\end{eqnarray}
The first and third lines of Eq. (\ref{eq2:Y}) can be found in Eq. (16.91) and Exercise 16.4.4 of Ref. \cite{Arfken2012}, respectively. The second line can be obtained from the equation $\Zjm\sqrt{j(j+1)}=i\phat\times\mathbf{L}Y_{jm}$. The result is reached by using the expression of the orbital angular momentum operator in momentum space $\mathbf{L}=i \nabla_{\pp}\times\pp$, the formula $\pp\times(\nabla_{\pp}\times\pp)=\frac{\nabla |\pp|^2}{2}-(\pp\cdot\nabla_{\pp})\pp$, and the gradient formula Eq. (16.94) in Ref. \onlinecite{Arfken2012}.

Figure \ref{fig:shell} illustrates Eq. (\ref{eq2:Y}). Table \ref{tab:t2} contains information about the $\Yjlm(\phat)$. Notably, we deduce from Eq. (\ref{eq2:Y}) and see in Fig. \ref{fig:shell} that, while $\Yjjm(\phat)$ is transverse, $\Yjjmm(\phat)$ and $\Yjjpm(\phat)$ are of mixed character, i.e., {\em neither transverse nor longitudinal}. 
\begin{table}[h!]
	\begin{ruledtabular}
\begin{center}\begin{tabular}{cccccc}
 & {$J^2$} & {$J_z$} & {$\Pi$}&{$L^2$}&Polarization\\
	\hline
 $\Yjjm(\phat)$ & $j(j+1)$ & $m$ & $(-1)^{j+1}$&$j(j+1)$&$\perp$\\
 $\Yjjmm(\phat)$ & $j(j+1)$ & $m$ & $(-1)^{j}$&$(j-1)j$&mixed\\
 $\Yjjpm(\phat)$ & $j(j+1)$ & $m$ & $(-1)^{j}$&$(j+1)(j+2)$&mixed\\
\end{tabular}\end{center}
	\end{ruledtabular}
	\caption{\label{tab:t2} Vector spherical harmonics: Polarization character and eigenvalues of $J^2$, $J_z$, $\Pi$, and $L^2$. $L^2$ is the orbital angular momentum squared. Both $j\ge 0$ and $m=-j\ldots j$ are integers. For $j=0$, $\mathbf{Y}_{0,1,0}=-\phat/(4\pi)$, $\mathbf{Y}_{0,0,0}=0$, and $\mathbf{Y}_{0,-1,0}$ is not defined.}
\end{table}

We can expand $\Jomega$ as:
	\begin{eqnarray}
		&&\Jomega=\\\sum_{j,m}\nonumber&&\tau_{j,j,m}^\omega\Yjjm(\phat)+\tau^\omega_{j,j-1,m}\Yjjmm(\phat)+\tau^\omega_{j,j+1,m}\Yjjpm(\phat).
	\end{eqnarray}
It follows from Eq. (\ref{eq2:Y}) that:
\begin{equation}
	\label{eq2:ac2}
	\begin{split}
		\bjm&=\tau^\omega_{j,j,m},\\
		\ajm&=-\Sjp\tau^\omega_{j,j-1,m}-\Sj\tau^\omega_{j,j+1,m},\\
		\cjm&=\Sj\tau^\omega_{j,j-1,m}-\Sjp\tau^\omega_{j,j+1,m}.
	\end{split}
\end{equation}

We invert Eq. (\ref{eq2:ac2}) to obtain:
\begin{equation}
	\label{eq2:ac2inv}
	\begin{split}
		\tau^\omega_{j,j,m}&=\bjm,\\
\tau^\omega_{j,j-1,m}&=-\Sjp\ajm+\Sj\cjm,\\
\tau^\omega_{j,j+1,m}&=-\Sj\ajm-\Sjp\cjm.
	\end{split}
\end{equation}

The current dependent integrals in $\rr$ space that define $\tau^\omega_{j,l,m}$ contain only the spherical Bessel function of order $l$. This can be shown using Eq. (\ref{eq2:ac2inv}) and the expressions for $\{\ajm,\bjm,\cjm\}$ in Eqs. (\ref{eq2:explicit}). For example, in the dipole case of Eq. (\ref{eq2:aexact}), the integrand that contains $j_0(kr)$ is only due to $\tau^\omega_{1,0,m}$, and the integrand that contains $j_2(kr)$ is only due to $\tau^\omega_{1,2,m}$. Both terms have zero out of shell components.

Let us assume that we want to determine $\tau^\omega_{j,j-1,m}$ and $\tau^\omega_{j,j+1,m}$ by measuring the electromagnetic radiation from the source at points $\rr$ such that $|\rr|>R$. As is clear from Eq. (\ref{eq2:ac2inv}), and can be visually appreciated in Fig. \ref{fig:shell}, the $\tau^\omega_{j,j\pm1,m}$ contain both transverse and longitudinal components. Their measurement involves the detection of both transverse and longitudinal fields generated by the current distribution. But, as previously discussed the longitudinal field due to the current distribution outside the source region is exactly canceled by that due to the charge distribution. Without access to the longitudinal degrees of freedom of $\Jomega$, it is impossible to separate the two $\tauj$ contributions to $\ajm$.

We see that there are splits of $\ajm$ that, while both parts remain free of out of shell components, they contain longitudinal components which render them non-separable. This is the case for a definition of the toroidal multipoles that has been recently given in Box 2 of Ref. \onlinecite{Papasimakis2016}. In App. D, we first show that this definition is different from the original definition in Refs. \onlinecite{Dubovik1974,Dubovik1990}. In the split of the $\ajm$ proposed in Box 2 of Ref. \onlinecite{Papasimakis2016}, the integrands contain entire spherical Bessel functions which prevent the appearance of out of shell components. We then show that each of the two parts contains longitudinal components that cannot be detected from outside the source region.

\section*{Discussion and conclusion \label{sec4}}
To finalize, we discuss some statements that are often found in the toroidal literature. 

It is often stated that the toroidal multipole family is a third family, independent of the electric and magnetic ones \cite{Kaelberer2010,Radescu2002,Fedotov2013,Papasimakis2016}. Originally, the authors of  Ref. \onlinecite{Dubovik1974} suggested in Sec. 4 to replace the $\{\ajm,\bjm,\cjm\}$ by $\{t^{\omega}_{jm},\bjm,\cjm\}$, where $t^{\omega}_{jm}$ are the coefficients of the toroidal part of $\ajm$. We have shown that the $t^{\omega}_{jm}$ cannot be separately determined, and can hence not constitute a third independent multipolar family.

It is also often stated that three multipolar families, including the toroidal one, are needed in order to expand a general charge-current distribution, while the radiation that they emit can be described with only two multipolar families \cite{Radescu2002,Savinov2014,Zhang2015}. Starting from Eq. (\ref{eq2:expand}), our analysis makes clear that, indeed, three multipolar families are needed to describe the source, and only two are needed to describe the fields radiated by it. It also makes clear that the third family is constituted by the longitudinal multipoles, and {\em not} the toroidal multipoles. The fact that the longitudinal field is zero outside the sources explains why the number of necessary multipolar families decreases by one when going from the sources to the fields. 

Finally, it is also stated in Refs. \onlinecite{Radescu2002,Papasimakis2016} that the toroidal multipoles can be separated using spectroscopic techniques. The separation is to be enabled by exploiting the extra frequency dependent weighting of the toroidal multipoles [$k^2=(\omega/c)^2$] with respect to the electric multipoles. It is straightforward to prove that, should this be actually possible, there would exist a doubly infinite number of independent multipolar families, and not just a third one. This unphysical outcome shows that the conjectured separability is not possible. In order to show this, let us assume that it is possible. The steps going from Eqs. (\ref{eq2:aexact}) to (\ref{eq2:tapprox}) reveal that the extra $k^2$ scaling comes from different terms in the series expansion of the spherical Bessel functions. We stopped at the second lowest order, but each of the infinite number of terms in the complete expansion has an extra $k^2$ weight with respect to its predecessor. Additionally, one can make a similar expansion of the magnetic multipoles $\bjm$ (see Sec. 4 in Ref. \onlinecite{FerCor2015b}). If the separate measurement of the terms with additional $k^2$ factors where possible, as conjectured in Box 2 of Ref. \onlinecite{Papasimakis2016}, there would be not just an extra independent multipolar family, but a doubly infinite number of them. The physical reality is that each of this doubly infinite number of {\em formally obtained} terms is not an independent degree of freedom. The multipolar electromagnetic transitions that determine the spectroscopic response of a system of charges and currents are completely characterized by using the two possible parities, electric and magnetic. The spectroscopic measurements of $\{\ajm,\bjm\}$ already contain all the available information. On the other hand, should the analysis of the spectroscopic data make use of models limited to the lowest order, the inclusion of the second lowest order terms can only improve it, and the use of exact expressions like Eq. (\ref{eq2:aexact}), Eq. (\ref{eq2:explicit}), and Eq. (20) in Ref. \onlinecite{FerCor2015b} will optimize it. 

In conclusion: Our analysis proves that the dynamic toroidal multipoles do not have an independent physical meaning with respect to their interaction with electromagnetic waves. Their formal meaning is clear, however: They are higher order terms of an expansion of the transverse multipolar coefficients of electric parity in the electromagnetic size of the source. 

\begin{acknowledgements}
We warmly thank Dr. Rasoul Alaee for numerous discussions and for his feedback on the manuscript. We also warmly thank Dr. M\'ario Silveirinha for pointing out that our analysis and results are valid independently of the shape of the volume confining the sources. I.F.-C. thanks Ms. Magda Felo for her help with the figures. The work was partially supported by the DFG within the SFB project 1173. S.N. also acknowledges support by the Karlsruhe School of Optics \& Photonics (KSOP).
\end{acknowledgements}
\section*{Author contributions}
I.F-C. did the analytical work that led to the results presented in the manuscript supported by S.N.. C. R. initiated and supervised the research. All authors discussed the results and contributed to the manuscript.
\section*{Competing financial interests}
The authors declare no competing financial interests.

\bibliographystyle{ifcbst}
\clearpage

\appendix
\section{Contributions to $\bjm$, $\ajm$ and $\cjm$\label{sec:acexp}}
Equation (\ref{eq:exact}) in \cite{FerCor2015b} is an exact expression for the $\{\ajm,\bjm,\cjm\}$ coefficients in terms of integrals in both momentum and coordinate space. With $\Qjm(\phat)$ standing for any of the $\{\Xjm(\phat),\Zjm(\phat),\Wjm(\phat)\}$ and $\qjm$ for any of the corresponding $\{\ajm,\bjm,\cjm\}$, Eq. (\ref{eq:exact}) in \cite{FerCor2015b} reads

\begin{equation}
	\begin{split}
		&\sqrt{\frac{\pi}{2}}\qjm=\\
	 &\sum_{\lbar\mbar}(-i)^{\lbar}{\intdOmegap \Qjm^\dagger(\phat) Y_{\lbar\mbar}(\phat)} \intdr \Jomegar Y^*_{\lbar\mbar}(\rhat)j_{\lbar}(kr),
\end{split}
\end{equation}
where $j_{l}(\cdot)$ are the spherical Bessel functions of the first kind.

As shown in \cite[App. \ref{sec:omte}]{FerCor2015b}, only terms with $\lbar=j$ contribute to the $\bjm$, while the $\ajm$ and $\cjm$ get contributions from both $\lbar=j-1$ and $\lbar=j+1$. Explicitly:
\begin{widetext}
{
\begin{equation}
	\label{eq2:explicit}
	\begin{split}
		&\sqrt{\frac{\pi}{2}}\bjm=(-i)^{j}\sum_{\mbar=-j}^{\mbar=j} \intdOmegap \Xjm^\dagger(\phat)Y_{j\mbar}(\phat)\intdr \Jomegar Y^*_{j\mbar}(\rhat)j_{j}(kr),\\
		&\sqrt{\frac{\pi}{2}}\ajm=\\
		&{(-i)^{j-1}}\sum_{\mbar=-(j-1)}^{\mbar=j-1} \intdOmegap \Zjm^\dagger(\phat)Y_{j-1\mbar}(\phat)\intdr \Jomegar Y^*_{j-1\mbar}(\rhat)j_{j-1}(kr)+\\
		&{(-i)^{j+1}}\sum_{\mbar=-(j+1)}^{\mbar=j+1} \intdOmegap \Zjm^\dagger(\phat)Y_{j+1\mbar}(\phat)\intdr \Jomegar Y^*_{j+1\mbar}(\rhat)j_{j+1}(kr),\\
		&\sqrt{\frac{\pi}{2}}\cjm=\\
		&{(-i)^{j-1}}\sum_{\mbar=-(j-1)}^{\mbar=j-1} \intdOmegap \Wjm^\dagger(\phat)Y_{j-1\mbar}(\phat)\intdr \Jomegar Y^*_{j-1\mbar}(\rhat)j_{j-1}(kr)+
		\\
		&{(-i)^{j+1}}\sum_{\mbar=-(j+1)}^{\mbar=j+1} \intdOmegap \Wjm^\dagger(\phat)Y_{j+1\mbar}(\phat)\intdr \Jomegar Y^*_{j+1\mbar}(\rhat)j_{j+1}(kr).
	\end{split}
\end{equation}
}
\end{widetext}

\section{Cancellation of longitudinal fields outside the source \label{sec:cancel}}
We show that the longitudinal field with $|\pp|=\omega/c$ is zero outside the source region.

We consider spatially confined monochromatic electric charge and current density distributions $\rho_\omega(\rr)$ and $\mathbf{J}_\omega(\rr)$ embedded in an isotropic and homogeneous medium with constant permittivity $\epsilon$ and permeability $\mu$. We assume them to be confined in space. In the Lorenz gauge, the scalar and vector potentials meet the following inhomogeneous wave equations: 
\begin{equation}
	\label{eq:first}
	\begin{split}
	\left(\nabla^2+(\omega/c)^2\right)\phi_\omega(\rr)&=\frac{-\rho_\omega(\rr)}{\epsilon},\\
	\left(\nabla^2+(\omega/c)^2\right)\Aomegar&=-\mu\mathbf{J}_\omega(\rr),	
	\end{split}
\end{equation}
where $c=1/\sqrt{\epsilon\mu}$.
Outside the source region, it can be shown that the spatial Fourier transforms of $\phi_\omega(\rr)$ and $\Aomegar$ are non-zero only for $|\pp|=\omega/c$ (see \cite[App. \ref{sec:onlyw}]{FerCor2015b}). With the help of \cite[Eq. (3.8)]{Devaney1974}, they are readily seen to be proportional to the Fourier components of the sources in the same spherical shell domain:

\begin{equation}
	\label{eq:phiA}
		\phiomega=\frac{\rhoomega}{4\pi\epsilon},\
		\Aomega=\frac{\mu\Jomega}{4\pi}.
\end{equation}

In coordinate space, the electric field as a function of the potentials is
\begin{equation}
	\mathbf{E}_\omega(\rr)=i\omega \mathbf{A}_\omega(\rr)-\nabla \phi_\omega (\rr),
\end{equation}
which, in momentum space ($\nabla\rightarrow i\pp$) reads
\begin{equation}
	\mathbf{E}_\omega(\phat)=i\omega\Aomega-i\pp \phiomega.
\end{equation}
The longitudinal electric field is hence 
\begin{equation}
 \label{eq:longE}
	\pp\cdot \mathbf{E}_\omega(\phat)=\pp\cdot\left(i\omega\Aomega-i\pp \phiomega\right).
\end{equation}
Using Eq. (\ref{eq:phiA}) and that $\pp$ is restricted to $\pp=\frac{\omega}{c}\phat$ we can write Eq. (\ref{eq:longE}) as
\begin{equation}
	\label{eq:aa}
	\begin{split}
		\frac{\omega}{c}\phat\cdot \mathbf{E}_\omega(\phat)&=\frac{\omega}{c}\phat\cdot\left(i\omega\frac{\mu\Jomega}{4\pi}-i\pp \frac{\rhoomega}{4\pi\epsilon}\right)\\
										 &=\frac{i\omega\mu}{4\pi}\left(\frac{\omega}{c}\phat\cdot\Jomega-\omega\rhoomega\right).
	\end{split}
\end{equation}

The term inside the brackets in Eq. (\ref{eq:aa}) is equal to zero because of the continuity equation in momentum space
\begin{equation}
	\nabla\cdot\Jomegar=i\omega\rho_\omega(\rr)\stackrel{\nabla\rightarrow i\pp}{\longrightarrow} i\pp\cdot\Jomega=i\omega\rhoomega\,
\end{equation}
 particularized at $|\pp|=\omega/c$.
 \begin{equation}
 i\frac{\omega}{c}\phat\cdot\Jomega=i\omega\rhoomega.
 \end{equation}

We conclude that, outside the source region, the longitudinal field with $|\pp|=\omega/c$ produced by the current density exactly cancels the one produced by the charge density.  Note that the result is gauge independent since it is a statement about the electric field.
\vspace{0.5cm}
\section{The split of electric and toroidal parts introduces out of shell components in both of them\label{app:split}}
In this appendix we show that the independent measurement of electric and toroidal parts is impossible.

Let us consider the expression of the exact frequency-dependent multipoles of electric parity $\ajm$ in Eq. (\ref{eq2:explicit}). The monochromatic current $\Jomegar$ appears in two different spatial integrals, 
\begin{equation}
		\label{eq:jm1}
\intdr \Jomegar Y^*_{j-1\mbar}(\rhat)j_{j-1}(kr),
\end{equation}
and
\begin{equation}
		\label{eq:jp1}
\intdr \Jomegar Y^*_{j+1\mbar}(\rhat)j_{j+1}(kr),
\end{equation}
where $k=\omega/c$, $j_l(\cdot)$ are spherical Bessel functions, $r=|\rr|$, $\rhat=\rr/|\rr|$, and $Y_{ln}(\cdot)$ are scalar spherical harmonics.

Let us now split Eq. (\ref{eq:jm1}) into two parts by means of the small argument expansion of $j_{j-1}(kr)$. We isolate the first term of the expansion, which is of order $(kr)^{j-1}$ and obtain:
\begin{equation}
		\label{eq:split}
		\begin{split}
				&\intdr \Jomegar Y^*_{j-1\mbar}(\rhat)j_{j-1}(kr)=\\
				&\intdr \Jomegar Y^*_{j-1\mbar}(\rhat) \frac{(kr)^{j-1}}{[2(j-1)+1]!!}+\\
				&\intdr \Jomegar Y^*_{j-1\mbar}(\rhat) \left\{j_{j-1}(kr)-\frac{(kr)^{j-1}}{[2(j-1)+1]!!}\right\},
		\end{split}
\end{equation}
where $n!!=n(n-2)(n-4)\ldots$ is the double factorial. 

As we will now show, this is the split that gives rise to the electric and toroidal parts in the original literature \cite{Dubovik1974,Dubovik1990}. The first term in Eq. (\ref{eq:split}) corresponds to the electric part, and the second term is contained in the toroidal part. The toroidal part also contains the whole contribution of the integrals involving $j_{j+1}(kr)$ in Eq. (\ref{eq:jp1}).

Let us now see this splitting in the original literature \cite{Dubovik1974,Dubovik1990}. We start from the definition of the time-dependent exact multipoles of electric parity, which can be written from Eqs. 20 and 24 in Ref. \onlinecite{Dubovik1974} (also from Eqs. 1.3 and 1.10 in Ref. \onlinecite{Radescu2002}):
\begin{widetext}
\begin{equation}
	\label{eq:a00}
	a_{jm}(k,t)=\intdr \left[\sqrt{\frac{j+1}{2j+1}}j_{j-1}(kr)\Yjjmm(\rhat)+\sqrt{\frac{j}{2j+1}}j_{j+1}(kr)\Yjjpm(\rhat)\right]^{\dagger}\Jrt,
\end{equation}
\end{widetext}
where $\mathbf{Y}_{jlm}(\cdot)$ are vector spherical harmonics. 

The split between electric and toroidal parts can be seen in Eq. 38 of Ref. \onlinecite{Dubovik1990}, and Eq. 4.6 of Ref. \onlinecite{Dubovik1990}: 
\begin{equation}
	\label{eq:a0}
	a_{jm}(k,t)=\partial_t Q_{jm}(0,t)+k^2T_{jm}(k,t),
\end{equation}
where the electric part is (Eq. 4.7 of Ref. \onlinecite{Dubovik1990})
\begin{equation}
	\label{eq:q0}
	\partial_t Q_{jm}(0,t)= \sqrt{4\pi j}\intdr r^{j-1}\Yjjmm(\rhat)^{\dagger}\Jrt,
\end{equation}
and $T_{jm}(k,t)$ is the toroidal part.

It can be seen from Eqs. (\ref{eq:a00}) to (\ref{eq:q0}) that the integrand that defines $T_{jm}(k,t)$ must contain the $j_{j+1}(kr)\Yjjpm(\rhat)$ contribution plus the $j_{j-1}(kr)\Yjjmm(\rhat)$ contribution except for the first term in the small argument expansion of $j_{j-1}(kr)$, which is of order $r^{j-1}$ and has been split up. This splitting corresponds to the one we have performed in Eq. (\ref{eq:split}). It causes the appearance of out of shell components in both electric and toroidal parts.

As mentioned in the main text and explained at the end of Sec. III in Ref. \onlinecite{FerCor2015b}, the spherical Bessel functions inside the spatial integrals act as a filter that completely reject the out of shell ($|\pp|\neq \omega/c$) components of the current. After the split, the term proportional to $(kr)^{j-1}$ by itself does not provide such rejection. This can be appreciated in the first line of Eq. (\ref{eq:split}): $r^{j-1}Y^*_{j-1\mbar}(\rhat)$ is a frequency independent function which cannot remove the $|\pp|\neq \omega/c$ components present in $\Jomegar$. Multiplication by a factor of $k^{j-1}$ does not change this. Since $\ajm$ are physically measurable quantities without out of shell components, the presence of $|\pp|\neq \omega/c$ components in the electric part implies their presence in the toroidal part with opposite sign, as it is obvious from Eq. (\ref{eq:split}).

The out of shell components do not couple to the electromagnetic field, and therefore preclude the independent physical measurement of the electric and toroidal parts.

\section{An alternative definition of the toroidal multipoles\label{appc}}
In this appendix we first show that the definition of toroidal multipoles recently given in \cite[Box 2]{Papasimakis2016} is different from the original definition in \cite{Dubovik1974,Dubovik1990}. Both definitions involve the split of the multipoles of electric parity into two parts, but the splits are different in \cite[Box 2]{Papasimakis2016} and \cite{Dubovik1974,Dubovik1990}. We demonstrate this by showing that the well known expression of the toroidal dipole in the limit of small source \cite[Eq. 2.11]{Dubovik1990} cannot be recovered from the definition of \cite[Box 2]{Papasimakis2016}. Some terms are missing. We then also show that the missing terms are contained in a coefficient which is explicitly excluded from the definition of toroidal multipoles in \cite[Box 2]{Papasimakis2016}. The derivations in this appendix recover one of the results from the main text: The toroidal dipole is just the next to leading order term in the small source expansion of the exact electric dipole. The difference is that here the result is obtained directly in coordinate ($\rr$) space, while Eq. (\ref{eq2:aexact}) from the main text was obtained in \cite{FerCor2015b} by first going to momentum ($\pp$) space, and then going back to $\rr$ space. Finally, we show that, in this alternative definition, both parts contain longitudinal terms, which render them non-separable.

We start by examining the definitions in \cite[Box 2]{Papasimakis2016}
\begin{equation}
\label{eq:box2}
	\begin{split}
		\mathbf{E}_{\text{sca}}(\rr)&=\frac{4\pi k^2}{c} \sum _{j,m} \left[Q_{jm}\mathbf{\Psi}_{jm}(\rr)+M_{jm}\mathbf{\Phi}_{jm}(\rr)+T_{jm}\mathbf{\Psi}_{jm}(\rr)\right],\\
	Q_{jm}&=\frac{c}{\sqrt{j(j+1)}}\intdr \rho_\omega(\rr) Y^*_{jm}(\rhat) \frac{d}{dr}\left[rj_j(kr)\right],\\
	M_{jm}&=\frac{1}{i\sqrt{j(j+1)}}\intdr \left[\nabla\cdot \left(\rr\times\Jomegar\right)\right] Y^*_{jm}(\rhat) j_j(kr),\\
	T_{jm}&=\frac{k}{\sqrt{j(j+1)}}\intdr \left[\rr\cdot\Jomegar\right]Y_{jm}^*(\rhat) j_j(kr),
	\end{split}
\end{equation}
where $\mathbf{E}_{\text{sca}}(\rr)$ is the field produced by the sources, and $\mathbf{\Psi}_{jm}(\rr)$ and $\mathbf{\Phi}_{jm}(\rr)$ are the multipolar fields of electric and magnetic parity, respectively. The $Q_{jm}$ are said to be charge excitations yielding electric multipoles, the $M_{jm}$ transverse (w.r.t $\rr$) current excitations yielding magnetic multipoles, and the $T_{jm}$ radial current excitations yielding toroidal multipoles.

We first note that $Q_{jm}$ and $T_{jm}$ are both multiplying the same multipolar field of electric parity $\mathbf{\Psi}_{jm}(\rr)$. This implies that, together, $Q_{jm}$ and $T_{jm}$ must completely determine the multipolar coefficients of electric parity. This can be readily checked by setting the magnetic currents to zero in Jackson's \cite[Eq. 9.167]{Jackson1998} expression for the exact multipoles of electric parity $a_E(j,m)$, namely:
\begin{equation}
\label{eq:aejm}
\begin{split}
	&a_E(j,m)=\frac{k^2}{i\sqrt{j(j+1)}}\times \\
	&\intdr Y_{jm}^*(\rhat) \left\{c\rho_\omega(\rr) \frac{d}{dr}\left[rj_j(kr)\right] + ik\left[\rr\cdot\Jomegar\right] j_j(kr)\right\}.
\end{split}
\end{equation}
It is clear from Eq. (\ref{eq:box2}) that $Q_{jm}$ corresponds to the first term of the sum in Eq. (\ref{eq:aejm}), and $iT_{jm}$ to the second term. The sum $Q_{jm}+iT_{jm}$ determines $a_E(j,m)$. The definition of $T_{jm}$ in Eq. (\ref{eq:box2}) involves a split of the $a_E(j,m)$ into two parts. We now show that it is a different split from the one in the original definition of toroidal multipoles \cite{Dubovik1974,Dubovik1990}. 

Let us use $iT_{1m}$
\begin{equation}
\label{eq:t1m}
iT_{1m}=\frac{ik}{\sqrt{2}}\intdr \left[\rr\cdot\Jomegar\right]Y_{1m}^*(\rhat) j_1(kr),
\end{equation}
to attempt to recover the toroidal dipole in the small source approximation \cite[Eq. 2.11]{Dubovik1990},
\begin{equation}
\label{eq:tdipole}
\tone=\intdr\frac{1}{10}\left\{\left[\rr\cdot\Jomegar\right]\rr-2r^2\Jomegar\right\},
\end{equation}
and which follows from the original definition (\cite[Eq. 39]{Dubovik1974}, \cite[Eq. 2.11]{Dubovik1990}).

We start by arranging the three components corresponding to $m=\{1,0,-1\}$ into a vector,
\begin{equation}
	\label{eq:tsph}
i\mathbf{T}_1^{\text{sph}}=i\begin{bmatrix}T_{11}\\T_{10}\\T_{1-1}\end{bmatrix}=\frac{ik}{\sqrt{2}}\intdr \left[\rr\cdot\Jomegar\right]\begin{bmatrix}Y_{11}^*\\Y_{10}^*\\Y_{1-1}^*\end{bmatrix}j_1(kr),
\end{equation}
and consider the correspondence between $Y_{1m}^*(\rhat)$ and $\rhat$ in the spherical basis
\begin{equation}
\label{eq:rhaty}
\rhat=\frac{\rr}{|\rr|}=\begin{bmatrix}\hat{r}_1\\\hat{r}_0\\\hat{r}_{-1}\end{bmatrix}=2\sqrt{\frac{\pi}{3}}\begin{bmatrix}-Y_{1-1}\\Y_{10}\\-Y_{11}\end{bmatrix}=2\sqrt{\frac{\pi}{3}}\begin{bmatrix}Y_{11}^*\\Y_{10}^*\\Y_{1-1}^*\end{bmatrix}.
\end{equation}
Equation (\ref{eq:tsph}) is a vector in the spherical vector basis except that, when both $\rr$ and $\Jomegar$ are expressed in the spherical vector basis, the term $\left[\rr\cdot\Jomegar\right]$ should be written $\left[\rr^{\dagger}\Jomegar\right]$, where $\rr^{\dagger}$ denotes the hermitian conjugate of the position vector $\rr$. This is due to the fact that, in the spherical vector basis, $\rr$ has complex components. In the Cartesian basis used in \cite[Box 2]{Papasimakis2016} and Eq. (\ref{eq:tdipole}), $\rr$ is real valued and the dot product $\left[\rr^{\dagger}\Jomegar\right]$ can be written as $\left[\rr\cdot\Jomegar\right]$. In this appendix we will use the Cartesian basis from now on. We can change $\mathbf{T}_1^{\text{sph}}$ to the Cartesian basis by multiplying $\mathbf{T}_1^{\text{sph}}$ itself, and the other vectors involved in the expression with the change of basis matrix 
\begin{equation}
\label{eq:C}
\begin{bmatrix}a_x\\a_y\\a_z\end{bmatrix}=
	\begin{bmatrix}
		\frac{-1}{\sqrt{2}}&0&\frac{1}{\sqrt{2}}\\
		\frac{-i}{\sqrt{2}}&0&\frac{-i}{\sqrt{2}}\\
		0&1&0\\
	\end{bmatrix}
\begin{bmatrix}a_{1}\\a_0\\a_{-1}\end{bmatrix}.
\end{equation}
After using Eq. (\ref{eq:rhaty}) and changing the basis, the Cartesian expression reads:
\begin{equation}
	\label{eq:t1c}
i\mathbf{T}_1=\frac{ik}{\sqrt{2}}\intdr \left[\rr\cdot\Jomegar\right]\rhat\sqrt{\frac{3}{\pi}} \frac{1}{2}j_1(kr).
\end{equation}

We now use $j_1(kr)\approx kr/3$ in the limit of small $kr$, to approximate Eq. (\ref{eq:t1c}) by 
\begin{equation}
\label{eq:t1mapp}
\begin{split}
i\mathbf{T}_1&\approx\frac{ik}{\sqrt{2}}\intdr \left[\rr\cdot\Jomegar\right]\rhat\sqrt{\frac{3}{\pi}} \frac{1}{2}\frac{kr}{3}\\
&=\frac{ik^2}{\sqrt{2}}\sqrt{\frac{3}{\pi}}\intdr \frac{\left[\rr\cdot\Jomegar\right]\rr}{6}.
\end{split}
\end{equation}
Equation (\ref{eq:t1mapp}) cannot reproduce Eq. (\ref{eq:tdipole}) because the $r^2\Jomegar$ terms present in Eq. (\ref{eq:tdipole}) are missing in Eq. (\ref{eq:t1mapp}). The definition of $T_{jm}$ in \cite[Box 2]{Papasimakis2016} is hence different from the original definitions of the toroidal multipoles in \cite{Dubovik1974,Dubovik1990}.

We now show that the missing terms are contained in $Q_{1m}$: They are the terms of second lowest order in the small source approximation of $Q_{1m}$. This reproduces our results from the main text.

We start with
\begin{equation}
Q_{1m}=\frac{c}{\sqrt{2}}\intdr \rho_\omega(\rr) Y^*_{1m}(\rhat) {\frac{d}{dr}\left[rj_1(kr)\right]},
\end{equation}
take the same steps as before regarding vectors and basis,
\begin{equation}
\mathbf{Q}_{1}=\frac{c}{\sqrt{2}}\intdr \rho_\omega(\rr) \rhat\sqrt{\frac{3}{\pi}} \frac{1}{2} \frac{d}{dr}\left[rj_1(kr)\right],
\end{equation}
and use the continuity equation $\rho_\omega(\rr)=\nabla\cdot\Jomegar/(ikc)$, to obtain:

\begin{equation}
\label{eq:q1m}
\mathbf{Q}_{1}=\frac{-i}{k\sqrt{2}}\intdr\left[ \nabla\cdot\Jomegar\right] \rhat\sqrt{\frac{3}{\pi}} \frac{1}{2} \shadedbox{\frac{d}{dr}\left[rj_1(kr)\right]}.
\end{equation}
We now consider the term in the shaded box, which, using the derivative of spherical Bessel functions\footnote{$\frac{d}{dx} j_l(x) = \frac{1}{2l+1}\left(lj_{l-1}(x)-(l+1)j_{l+1}(x)\right)$} can be written as 
\begin{equation}
\label{eq:aiai}
{\frac{d}{dr}\left[rj_1(kr)\right]}=j_1(kr)+\left\{\frac{kr}{3}\left[j_0(kr)-2j_2(kr)\right]\right\}.
\end{equation}
We now take terms up to order $(kr)^3$ in the small argument approximation of the spherical Bessel functions of Eq. (\ref{eq:aiai}): 
\begin{equation}
	\begin{split}
		j_0(kr)&\approx 1-\frac{(kr)^2}{6},\\
		j_1(kr)&\approx \frac{kr}{3}\left(1-\frac{(kr)^2}{10}\right),\\
		j_2(kr)&\approx \frac{(kr)^2}{15},\\
	\end{split}
\end{equation}
to obtain the approximate expression
\begin{equation}
\label{eq:appshade}
\begin{split}
&{\frac{d}{dr}\left[rj_1(kr)\right]}\approx\\& \frac{kr}{3}\left(1-\frac{(kr)^2}{10}\right)+\left\{\frac{kr}{3}\left[1-\frac{(kr)^2}{6}-2\frac{(kr)^2}{15}\right]\right\}.
\end{split}
\end{equation}
We now plug Eq. (\ref{eq:appshade}) into the right hand side of Eq. (\ref{eq:q1m}) 
\begin{equation}
	\label{eq:fr1}
	\begin{split}
		\mathbf{Q}_1 &\approx \frac{-i}{k\sqrt{2}}\sqrt{\frac{3}{\pi}}\frac{1}{2}\intdr\left[ \nabla\cdot\Jomegar\right]\rhat \frac{2(kr)}{3}+\\
	&\frac{-i}{k\sqrt{2}}\sqrt{\frac{3}{\pi}}\frac{1}{2}\intdr\left[ \nabla\cdot\Jomegar\right]\rhat  \frac{(kr)^3}{3}\left(-\frac{1}{10}-\frac{1}{6}-\frac{2}{15}\right),
	\end{split}
\end{equation}
and reduce it to
\begin{equation}
	\label{eq:fr2}
	\begin{split}
		\mathbf{Q}_1 &\approx \frac{-i}{\sqrt{6\pi}}\intdr \left[\nabla\cdot\Jomegar\right]\rr +\\
						   &\shadedbox{i\frac{k^2}{\sqrt{2}}\sqrt{\frac{3}{\pi}}\intdr \frac{\left[ \nabla\cdot\Jomegar\right] r^2\rr}{15}},
	\end{split}
\end{equation}
Using steps similar to those in \cite[Eq. 9.14]{Jackson1998}, the first line of Eq. (\ref{eq:fr2}) can be readily brought to the familiar form of the approximate electric dipole for small sources:
\begin{equation}
	\frac{i}{\sqrt{6\pi}}\intdr \Jomegar.
\end{equation}
This lowest order term is precisely the one separated by Dubovik \cite{Dubovik1974,Dubovik1990} in his split of the exact multipoles of electric parity between what is referred to as the ``electric'' part, which is this term in \cite[Eq. 9.170]{Jackson1998}, and the toroidal part, which are the higher order terms.

Incidentally, reversing the use of the continuity equation in the first line of Eq. (\ref{eq:fr2}) recovers the first line in \cite[Eq. 9.170]{Jackson1998}.

We are now interested in the next to leading order terms of $\mathbf{Q}_1$, which are shaded in the second line of Eq. (\ref{eq:fr2}). We now show that, when summed to Eq. (\ref{eq:t1mapp}), the toroidal dipole is recovered. The key step is to use the divergence theorem in the integration by parts of Eq. (\ref{eq:fr2}). To that end we first consider the following steps, where $n$ and $k$ run over $\{1,2,3\}$ and $\delta_{nk}$ is the Kronecker delta:
\begin{equation}
\label{eq:div}
\begin{split}
&\nabla\cdot\left(\mathbf{J} r^2 r_k\right)=
\sum_n \left[\left(\partial_n J_n\right)r^2r_k+J_n\partial_n(r^2r_k)\right]=\\
&\sum_n \left[\left(\partial_n J_n\right)r^2r_k+J_n(2r_nr_k+r^2\delta_{nk})\right]=\\
&\left(\nabla\cdot\mathbf{J}\right)r^2r_k+2\left(\rr\cdot\mathbf{J}\right)r_k+J_kr^2.
\end{split}
\end{equation}
Since $\Jomegar$ is, by assumption, bounded in space, the divergence theorem can be used to show that 
\begin{equation}
	\intdr \nabla\cdot\left[\Jomegar r^2 r_k\right] =0 \text{ for all } k.
\end{equation}
The divergence theorem \cite[p. 36]{Jackson1998}
\begin{equation}
	\label{eq:divt}
\int_V\ d^3\rr\ \nabla\cdot\mathbf{A} = \oint_S\ \mathbf{A}\cdot \rhat\ dS
\end{equation}
relates the integral of the divergence of any well-behaved vector field $\mathbf{A}$ over a volume $V$ to the flux through the surface boundary of $V$. For our purposes we set $\mathbf{A}= \Jomegar r^2 r_k$, and choose $V$ as a spherical volume enclosing the sources, so that $\Jomegar r^2 r_k$ is zero on its surface boundary. Then, the right hand side of Eq. (\ref{eq:divt}) vanishes.

Therefore, using Eq. (\ref{eq:div}) for $k=\{1,2,3\}$ we obtain
\begin{equation}
\label{eq:divnice}
\begin{split}
	\intdr & \left[\nabla\cdot\Jomegar\right]r^2\rr =\\ &- 2\intdr \left[\rr\cdot\Jomegar\right]\rr-\intdr r^2\Jomegar,
\end{split}
\end{equation}
whose left hand side appears in Eq. (\ref{eq:fr2}). After substituting Eq. (\ref{eq:divnice}) into Eq. (\ref{eq:fr2}) we get:
\begin{equation}
\label{eq:q1mapp2}
i\frac{k^2}{\sqrt{2}}\sqrt{\frac{3}{\pi}}\intdr \frac{1}{15} \left\{-2\left[\rr\cdot\Jomegar\right]\rr-r^2\Jomegar\right\}.
\end{equation}
Let us recall that Eq. (\ref{eq:q1mapp2}) are the next to leading order terms in the small source expansion of $Q_{1m}$. After summing them to Eq. (\ref{eq:t1mapp})
\begin{equation}
\begin{split}
&i\frac{k^2}{\sqrt{2}}\sqrt{\frac{3}{\pi}}\intdr \left(-\frac{2}{15}+\frac{1}{6}\right)\left[\rr\cdot\Jomegar\right]\rr-\frac{1}{15}r^2\Jomegar=\\
&i\frac{k^2}{\sqrt{6\pi}}\intdr \frac{1}{10}\left\{\left[\rr\cdot\Jomegar\right]\rr-2r^2\Jomegar\right\},
\end{split}
\end{equation}
we recover the exact form of the toroidal dipole in Eq. (\ref{eq:tdipole}). 

This derivation also recovers a result from the main text: The toroidal multipoles are just higher order terms in the small source approximation of the exact multipoles of electric parity.

To conclude, we note that, according to Eq. (\ref{eq:q1m}), which contains $\nabla\cdot\Jomegar$, $\mathbf{Q}_1$ depends on the longitudinal degrees of freedom of $\Jomegar$. This follows from the $\rr$ domain to $\pp$ domain correspondence
\begin{equation}
\nabla\cdot\Jomegar\rightarrow i\pp\cdot\Jomegagen,
\end{equation}
which makes clear that the $\cjm$ coefficients multiplying the longitudinal multipoles contribute to $\mathbf{Q}_1$ [see Eqs. (\ref{eq2:expand}) and (\ref{eq2:xzw})]. 

Since the sum $Q_{jm}^\omega+iT_{jm}^\omega$ is equivalent to the fully transverse $\ajm$, it then follows that the longitudinal dependence introduced by $\mathbf{Q}_1$ must be canceled by $i\mathbf{T}_1$. The split introduces longitudinal components in both parts, which renders them non-separable according to the discussion in the main text (see the section entitled ``Splits without out of shell components'').
 \end{document}